\documentclass[conference]{IEEEtran}
\usepackage{xcolor,colortbl}
\usepackage{tabstackengine}
\usepackage{multirow}
\usepackage{amsmath}
\usepackage{tabularx}
\usepackage{booktabs}
\usepackage{hyperref}
\usepackage{graphicx}
\usepackage{hhline}
\usepackage{comment}
\usepackage{makecell}
\usepackage[framemethod=tikz]{mdframed}

\providecommand{\keywords}[1]{\textbf{\textit{Keywords---}} #1}

\usepackage{geometry}
\definecolor{confmcolor}{HTML}{08306b}
\definecolor{blockbody}{cmyk}{0, 0, 0, 0.1}
\mdfdefinestyle{simple3}
{
linewidth=0.5pt,
roundcorner = 10pt,
backgroundcolor = blockbody,
linecolor = white,
innertopmargin = \topskip,
splittopskip = \topskip
}

\newcommand\confmat[3]{\multicolumn{1}{|>{\columncolor{confmcolor!#1}}c|}{\tabularCenterstack{c}{\color{#3}#1\% \\ \color{#3}\small{(#2)}}}}
\newcommand\confmatl[3]{\multicolumn{1}{>{\columncolor{confmcolor!#1}}c|}{\tabularCenterstack{c}{\color{#3}#1\% \\ \color{#3}\small{(#2)}}}}
\newcommand\confmatr[3]{\multicolumn{1}{>{\columncolor{confmcolor!#1}}c|}{\tabularCenterstack{c}{\color{#3}#1\% \\ \color{#3}\small{(#2)}}}}

\begin{document}

\title{Code Detection for Hardware Acceleration Using Large Language Models}
\date{}

\author{

\IEEEauthorblockN{Pablo Antonio Martínez}
\IEEEauthorblockA{\textit{Computer Engineering Department} \\
\textit{University of Murcia}\\
Murcia, Spain \\
pabloantonio.martinezs@um.es}

\and

\IEEEauthorblockN{Gregorio Bernabé}
\IEEEauthorblockA{\textit{Computer Engineering Department} \\
\textit{University of Murcia}\\
Murcia, Spain \\
gbernabe@um.es}

\and

\IEEEauthorblockN{José Manuel García}
\IEEEauthorblockA{\textit{Computer Engineering Department} \\
\textit{University of Murcia}\\
Murcia, Spain \\
jmgarcia@um.es}

}
\maketitle

\thispagestyle{plain}
\pagestyle{plain}
        
\begin{abstract}
Large language models (LLMs) have been massively applied to many tasks, often surpassing state-of-the-art approaches.
While their effectiveness in code generation has been extensively studied (e.g., AlphaCode), their potential for code detection remains unexplored.

This work presents the first analysis of code detection using LLMs.
Our study examines essential kernels, including matrix multiplication, convolution, and fast-fourier transform, implemented in C/C++.
We propose both a preliminary, naive prompt and a novel prompting strategy for code detection.

Results reveal that conventional prompting achieves great precision but poor accuracy (68.8\%, 22.3\%, and 79.2\% for GEMM, convolution, and FFT, respectively) due to a high number of false positives.
Our novel prompting strategy substantially reduces false positives, resulting in excellent overall accuracy (91.1\%, 97.9\%, and 99.7\%, respectively).
These results pose a considerable challenge to existing state-of-the-art code detection methods.

\end{abstract}


\section{Introduction and Motivation}
In recent years, transformer-based models have superseded state-of-the-art neural network models for all kinds of tasks, like computer vision~\cite{bao2022beit}, natural language processing (NLP)~\cite{ouyang2022training}, speech recognition~\cite{radford2022robust} or translation~\cite{bart2020}.
One of the most impressives models based on the transformer architecture~\cite{attention} are large language models (LLMs).
A LLM is no more than a language model trained with significantly larger training data than traditional language models.
In essence, LLMs are probability distributions over sequences of words.
That is to say, language models aim to understand and generate text, just like humans do.

LLMs are provoking impressive impact in diverse deep learning fields, thanks to their impressive abilities to perform diverse tasks~\cite{chagptjack, llmsurvey}.
The extensive list of applications where LLMs works well implies that there may still be unexplored possibilities awaiting discovery.
In particular, even though LLMs are proven to work for code synthesis~\cite{llmsurvey}, their evaluation in code detection remains unexplored.
Code detection involves analyzing programs to identify the algorithms they contain, which is crucial in numerous areas of programming language research.
A notable application of code detection is to match and replace handwritten code with optimized libraries.

Specialized hardware accelerators provide massive performance and energy efficiency improvements over traditional microprocessors~\cite{accelerators}.
However, there is a lot of code that is already written for CPUs, so executing it on accelerators is not trivial.
Detecting parts of acceleratable code and replacing it with appropriate API calls is a novel approach to overcome these issues. %
This technique has many advantages, like improving the performance of a hardcoded implementation or offloading compute-heavy tasks to accelerators automatically.
Many works focused on this topic~\cite{idl,kfr,facc,atc}, which rely either on constrained-based pattern matching or neural network-based code detection.
Those code detection techniques are generally either brittle, unable to match complex code, or inefficient in recognizing code patterns.
On the contrary, LLMs are trained with a huge corpus of source code, so code detection is a field where LLMs can potentially outstand.

Despite LLM powers, it is still difficult to exploit their full capabilities because they are very sensitive to the prompt.
LLMs can suffer from different kinds of faults, like hallucination~\cite{hallucination}, which has motivated the emergence of a new discipline called prompt engineering.
Prompt engineering aims to build prompts for maximizing the performance of LLMs.
However, depending on the specific application, different prompt engineering methods shall be used because not all work well for every situation.

In this paper, we propose a novel methodology that leverages LLMs to perform code detection.
More specifically, we investigate the use of the GPT-3.5 platform and OpenAI's API for code detection tasks.
To evaluate its performance, we build a benchmark suite consisting of significant and computationally intensive algorithms from computer science.
Our evaluation demonstrates that GPT-3.5 successfully detects a 93\% of matrix multiplication (GEMM), 100\% convolution, and 100\% fast-Fourier transform (FFT) programs.
Furthermore, we evaluate GPT-3.5 using a benchmark containing diverse program implementations that do not include the target algorithms.
This evaluation reveals a false positive rate of 43\%, 80\%, and 21\% for GEMM, convolution, and FFT, respectively.
We investigate the causes of this issue and propose a novel prompting strategy for code detection using prompt engineering.
Using our novel proposal we significantly reduce the false positive rates to 5.8\% for GEMM, 1.6\% for convolution, and 0.0\% for FFT, significantly improving the accuracy.
Our results affirm the viability of LLMs, specifically GPT-3.5, for code detection tasks and their potential integration with existing compilation techniques to replace handwritten code with accelerated implementations.

This paper makes the following contributions:
\begin{itemize}
\item We present the first analysis on code detection using large language models.
\item We propose a novel prompting strategy for code detection that drastically improves LLM performance at code detection tasks.
\item We evaluate GPT-3.5 in code detection against matrix multiplication, convolution and fast-fourier-transform programs, as well as other several programs to evaluate GPT-3.5 false positives rates.
\end{itemize}

The rest of this paper is organized as follows.
Section~\ref{sec:s2} presents the background on prompt engineering and large language models, as well as code detection techniques and their applications for hardware acceleration.
We present our methodology to use GPT-3.5 API for code detection in Section~\ref{sec:s3}.
Section~\ref{sec:s4} shows our prompt engineering work, divided into two approaches, the second being a novel prompting technique for code detection.
In Section ~\ref{sec:s5} we evaluate the accuracy of each prompting technique, showing that GPT-3.5 is indeed a code detection tool for hardware acceleration.
Finally, Section~\ref{sec:s6} concludes the work and gives hints and suggestions for future work.

\section{Background and related work} \label{sec:s2}

\subsection{Large Language Models (LLMs)}
Large Language Models (LLMs) refer to a family of language models, based on the transformer architecture~\cite{attention} containing billions of parameters that are trained on massive datasets containing text~\cite{llmsurvey}.
Generative Pre-trained Transformer 3 (GPT-3) is a 175 billion parameter LLM released in 2020 by OpenAI~\cite{gpt3}.
Two years later, OpenAI released a new subclass, GPT-3.5, which is the base of the popular ChatGPT, a chatbot fine-tuned for conversations that have caused a massive shock in society~\cite{chatgptopinion,microsoftgpt4}.
OpenAI's last release is GPT-4~\cite{gpt4}, a new version in the GPT series, which has shown impressive results in many tasks, leading us to think that we are approaching artificial general intelligence (AGI)~\cite{microsoftgpt4}.
Other notable LLMs include Pathways Language Model (PaLM)~\cite{palm} (Google) or LLaMA~\cite{llama} (Meta).

Applications where LLM exceeds are uncountable.
From coding to mathematical abilities, LLMs' power focus on natural language tasks.
One particular aspect of LLM is their capability to show abilities that are not present in small models but arise in large models~\cite{abilities}.
Generally speaking, we can organize these abilities into three domains: in-context learning (ICL), instruction following and step-by-step reasoning (e.g., chain-of-thought)~\cite{abilities,llmsurvey}.
ICL consists of prompting LLMs with a question in natural language description along with several demonstrations and a test query~\cite{llmsurvey}.
In instruction following, tasks are described in the form of instructions.
With instruction tuning, LLMs can follow the task instructions for new tasks without using explicit examples, thus improving their ability to generalize.
Lastly, step-by-step prompting includes several intermediate reasoning steps, which LLMs can use to perform complex reasoning.

\subsection{Prompt Engineering}
Language models are designed to imitate and predict the next token given a sequence of tokens as input~\cite{firstlm}, which explains why LLMs are extremely sensitive to the prompt.
Thus, it is crucial to understand how to provide good prompts to LLMs in order to achieve good quality outputs.
In this sense, prompt engineering covers a set of techniques to improve the communication between humans and LLMs.
Prompts are instructions given to an LLM to specify the quality and quantity of the output, enforce rules in the output, etcetera.
Prompting can actually also be considered as a form of programming a LLM, since we are giving precise instructions of how and what to do~\cite{promptpatterncatalog}.
In that respect, recent works have proposed new languages to aid programming of LLMs~\cite{lqml,guidance}.

Zero-shot prompting is the most straightforward prompting technique, where the prompt contains only instructions describing the task.
On the contrary, few-shot prompting consists in providing additional examples or demonstrations to the model.
This technique can improve LLM performance on complex tasks, where zero-shot prompting is not enough.

Like humans, LLMs are thought to be capable of performing reasoning~\cite{mialon2023augmented}.
In this line, one critical prompt engineering technique to improve reasoning is called chain-of-thought (CoT)~\cite{cot}.
With CoT, the prompt is spread out across a larger sequence of tokens. %
Instead of prompting the model with just the question, CoT prompts include examples of chain-of-thought sequences of the task.
In this sense, CoT is an instance of few-shot prompting, proposing a simple solution by modifying the answers in few-shot examples to step-by-step answers~\cite{cot2}.

Another relevant technique is called self-consistency~\cite{selfconsistency}.
Intuitively, it is based on the idea that we, as humans, think in different ways.
In tasks requiring reasoning, it is natural to have several ways to attack the problem.
Self-consistency consists in prompting the model using CoT multiple times to sample a diverse set of reasoning paths.
Afterward, it selects the most consistent answer by marginalizing all the available answers in the answer set.
This method has proved to be effective in several scenarios~\cite{selfconsistency}.

A more complex technique, based on the previous approaches of CoT and self-consistency is tree-of-thoughts (ToT)~\cite{tot}.
This paradigm allows LLMs to explore multiple reasoning paths over thoughts.
In this context, a thought is considered a part of the final solution.
ToT prompting involves defining four key aspects:  how to decompose the intermediate process into thoughts, how to generate potential thoughts from each state, how to evaluate each state and what search algorithm to use.
This technique has shown promising results in solving complex tasks such as Game of 24, Creative Writing, and Crosswords~\cite{tot}.

A simpler, yet crucial aspect of prompting is controlling the format of the output.
For example, it is useful to use delimiters to demarcate sections of text to be treated differently~\cite{gptbestpractices}.
Other techniques for controlling the output format are specifying the desired length of the output or the format and order in which each part of the output should be presented.

\subsection{Programming Hardware Accelerators}

Specialized hardware accelerators provide massive performance and energy efficiency improvements over traditional microprocessors~\cite{accelerators}, necessary to overcome the challenges of increasingly complex and compute-demanding applications.
Instead of using one device (CPU) for everything, accelerators are specialized for a given domain. 
The benefit of microprocessors' generality is their ease of adaption, but at the same time, it is their source of inefficiency.
However, accelerators are highly diverse~\cite{surveyacc}, which makes programming particularly hard.

To write new code for accelerators, the typical approach is writing software in the programming language specifically designed for that accelerator.
For existing code (e.g., a program written in a general-purpose language) there is a better alternative than rewriting the program using the accelerator-specific language.
This alternative involves replacing parts of the code with calls to the accelerator API~\cite{idl,kfr}.
By utilizing libraries that target the accelerator API, this approaches allows the compilation of old code, effectively using the accelerator without the need for extensive code rewriting.

To replace acceleratable parts of code with calls to an optimized library, the compiler must accomplish two distinct tasks.
First, the compiler must detect parts of the code that are suitable for replacement.
Second, the compiler must find the mappings between the variables in the original program and the variables in the library.
In this work, we focus on exploiting LLMs to achieve the first task, code detection, which discovers the acceleratable parts of the code that are candidates for replacement.

\subsection{Code Detection}
Code detection approaches in the literature can be divided into two categories:
\begin{itemize}
\item \textbf{Constraint-based matching}:
Also known as pattern-based matching, consists in finding constraints and patterns in the code that can be matched into a previously defined set of constraints~\cite{proganalysisbook}.
Idiom Description Language (IDL)~\cite{idl} proposed a description language that allows the user to define constraints for detecting particular idioms.
Idioms are later translated into a set of constraints over the LLVM IR~\cite{llvm}, which are used by the compiler to match the corresponding code.
In~\cite{mlirrewriting}, authors also focus on idiom matching and rewriting, but the idiom specification aims to be easier to understand.
Besides, idioms are translated into MLIR~\cite{mlir} dialects rather than LLVM IR.
Similarly, a pattern constraint-based approach is used in KernelFaRer~\cite{kfr}, where authors focus primarily on detecting matrix-multiplication (GEMM) and SYR2K kernels.
Here, constraints are hardcoded inside the compiler, and the user can not modify them.
Constraint-based matching is, however, very brittle and generally unable to match complex code structures~\cite{atc}.

\item \textbf{Neural embeddings}:
A neural network is trained to detect the code.
This is a more modern approach to code detection~\cite{facc,atc}, which uses neural embeddings to detect acceleratable regions.
In~\cite{facc} and~\cite{atc}, the neural classifier is based on ProGraML~\cite{programl}, and it is trained using the OJClone dataset~\cite{ojclone}, containing 105 classes of different algorithms, each implemented in different ways.
New algorithms not present in the dataset can be easily identified by adding examples of that particular algorithm to the dataset, which will make the neural classifier learn how to identify them.
The downside of current neural embedding proposals is their low accuracy.
They perform well to guide the search of a given algorithm but although they can match more complex code structures, they are less reliable compared to constraint-based matching.
\end{itemize}

Above mentioned techniques aid to find and categorize sections of code, but they do not provide any guarantee that those sections are correctly identified, which can potentially lead to faulty compilation.
Therefore, some verification is typically coupled with code detection to prove that the code was identified correctly and works as expected.
Besides formal verification techniques, other approaches rely on comparing the output of the code with the output of a valid program (input-output validation)~\cite{facc,atc}.
Despite the efforts to ensure the correctness of code, programmer sign-off is ultimately required.

\section{Using GPT-3.5 to Analyze Code} \label{sec:s3}
\subsection{Using the API}
At the time of writing, OpenAI's models (GPT-3.5, GPT-4) and Claude are the most competitive general-purpose LLMs~\cite{llmranking}~\footnote{Ranking available at: \url{https://chat.lmsys.org/?leaderboard}}.
Another LLM to consider is AlphaCode~\cite{alphacode}, a model specifically trained to generate code.
However, it is not publicly available, and it is limited to code generation, not code detection.
While Claude and GPT-4 are in a limited beta (they are only accessible to those who have been granted access), GPT-3.5 (the base model for ChatGPT) is publicly available~\cite{gptmodels}.
In this work, we use the \texttt{gpt-3.5-turbo-16k} model (the most capable GPT-3.5 model, according to OpenAI~\cite{gptmodels}) with the OpenAI API.
The \texttt{16k} suffix simply indicates that this model has a 16k context window, in contrast to the standard GPT-3.5 model which has a context window of 4k tokens.
Having a larger context window allows to analyze larger codes that otherwise would not fit in 4k or fewer tokens.
We believe that the proposals shown in this work are applicable to any other LLM and that results shall be similar.

We designed a simple wrapper around the OpenAI API written in Python.
It is is responsible of reading the source code, prompting the \texttt{gpt-3.5-turbo-16k} model with the code and parsing the output.
As we will see in Section~\ref{sec:s4}, our prompt includes formatting instructions.
However, sometimes GPT-3.5 produces outputs that do not match exactly the specified output format.
This motivates the need of a simple parser (detailed in Section~\ref{sec:rules}) which allows our Python program to interpret the output (e.g., to understand when a program is correctly classified and when it is not).
The API can be configured with parameters such as the role or the temperature and nucleus sampling parameters.
The role parameter indicates how the model should behave (acting as in a given role, such as system, user, assistant, or function).
Temperature and nucleus sampling (\texttt{top\_p} parameter) allows to control the sampling of the model.
Both are ways to control the sampling, so it is recommended to alter only one of those, but not both.
Intuitively, higher temperature values will make the output more random, while lower values will make it more deterministic.

\subsection{The Dataset} \label{sec:dataset}
In this work, we focus on detecting C/C++ codes, so all codes in our dataset are implemented in one of those languages.
We identify the three most relevant kernels in code detection studies.
GEMM is clearly the one that gets most of the attention~\cite{idl,kfr,atc}.
Fast-fourier transform (FFT) and convolution are also studied in the literature~\cite{facc,atc}.
To create the dataset, we explored GitHub C/C++ code performing any of those three kernels.
To better understand the quality of GPT-3.5 code detection, we focused on finding different implementations of each algorithm.

\subsubsection{GEMM}
We identified and gathered 7 classes of matrix multiplications:

\begin{itemize}
\item {\em Naive:} Implementations with the traditional 3-loop structure.
\item {\em Unrolled:} Implementation with unrolled loops.
\item {\em Function Calls:} Implementations dividing the compute into different function calls.
\item {\em Tiled:} Tiled implementations.
\item {\em Goto:} Implementations using the Goto algorithm~\cite{goto}.
\item {\em Strassen:} Implementations using the Strassen algorithm~\cite{strassen}.
\item {\em Intrinsics:} Implementations using Intel intrinsics.
\end{itemize}

\subsubsection{Convolution}
We found 3 different implementations:
\begin{itemize}
\item {\em Winograd:} The Winograd algorithm.
\item {\em Direct:} The direct convolution algorithm.
\item {\em im2col+gemm:} Uses a method called \textit{im2col} to compute the convolution using GEMM (e.g., like the Caffe framework~\cite{caffe}).
\end{itemize}

\subsubsection{FFT}
We retrieved 3 different implementations:
\begin{itemize}
\item {\em DFT:} Discrete fourier transform implementations.
\item {\em Radix-2:} Computes the DFTs of the even-indexed and the odd-indexed inputs separately and then combines both.
\item {\em Recursive:} Recursive implementations.
\end{itemize}

\begin{table}[t]
\centering
\caption{Source code dataset description (available at~\cite{zenododataset})} \label{tab:dataset}
\begin{tabular}{llll}
\toprule
                                                                                                  & Code         & Number of Codes  & Total LoC  \\ \midrule
\multirow{3}{*}{\rotatebox[origin=c]{90}{\parbox[c]{1cm}{\centering \textbf{Real Code}}}}         & GEMM         & 128              & 11.3k      \\
                                                                                                  & Convolution  & 15               & 2.8k       \\
                                                                                                  & FFT          & 15               & 1.8k       \\
                                                                                                  & Total        & 158              & 16.0k      \\ \midrule
\multirow{3}{*}{\rotatebox[origin=c]{90}{\parbox[c]{1.5cm}{\centering \textbf{False Positives}}}} & Parboil      & 10               & 1.4k       \\
                                                                                                  & Caffe        & 182              & 42.3k      \\
                                                                                                  & ACOTSP       & 13               & 2.8k       \\
                                                                                                  & cpufetch     & 22               & 5.7k       \\
                                                                                                  & Total        & 227              & 52.3k      \\
\bottomrule
\end{tabular}
\end{table}

\subsubsection{False Positives}
Also, we are interested in measuring the exposure of the model to false positives. %
Therefore, we also included programs not explicitly containing any of the three aforementioned algorithms.
We differentiate this part of the dataset between mainstream and not mainstream code.
For the mainstream code, we included the Parboil benchmark~\cite{parboil}, a set of applications for benchmarking the performance and throughput of processors, and Caffe~\cite{caffe}, a deep learning framework.
Those programs are somewhat popular and it is easy to find similar implementations of those applications in the wild.
Besides, the Parboil benchmark contains a matrix-vector multiplication, which can be useful to understand the sensibility of the model, since it is a kernel very similar to matrix multiplication.
For the non-mainstream code, we included cpufetch~\cite{cpufetch}, a program that gathers CPU architecture information, and an Ant Colony Optimization (ACO) implementation~\cite{acotspmf}.
Furthermore, for those codes that unintentionally contained GEMM, convolution or FFT, we removed them from the dataset to make sure that those codes do not contain such algorithms.
A description of the dataset is shown in Table~\ref{tab:dataset}.

\subsection{Feeding GPT-3.5 with Code}
To feed the model with code, we simply copy and paste the code into the prompt.
In other words, we use crude code straight into the model.
It is worth noting that LLMs have an input token limit, meaning that they can only process inputs smaller than their limit.
If the code is larger than the token limit, we identify two ways of processing the input.
First, decreasing the token count of the code.
The idea is to reduce the token count without changing the code semantics (e.g., replacing spaces with tabs).
Removing comments, removing dead code, or reducing the length of variables and function names are also viable, but they might hurt the model's ability to reason about code.
Second, partition the code into smaller parts.
For example, partitioning the original program into $n$ partitions.
In our case, however, none of the codes surpassed the 16k token limit, so we can safely feed GPT-3.5 with code without additional modifications.

\section{Prompt Engineering} \label{sec:s4}
One of the most common tips for good prompting is to start with a simple prompt and then iterate over more complex and complete prompts.
Here we describe our prompt engineering process in which we started with a very first prompt and a second version of the prompt that is aimed to improve it.

\subsection{First Prompt}
\begin{figure}[ht!]
\caption{First prompt.} \label{fig:prompt1}
\begin{mdframed}[style=simple3]
I want to know if the code below contains any function performing a *\textit{algorithm}*. Please ignore functions whose definition is not visible.\\
\\

Desired format:\\
Yes: {function name} (if there is a function).\\
No (if there is no function)\\
\\

Code:``````\\
*\textit{the actual code}* \\
''''''
\end{mdframed}
\end{figure}

The first prompt is detailed in Figure~\ref{fig:prompt1}.
In this prompt, the keyword *\textit{algorithm}* is replaced by the specific algorithm we are looking for.
That is, *\textit{algorithm}* may take the value of ``matrix multiplication (GEMM)'', ``convolution'' or ``fast Fourier transform (FFT)''.
Also, the keyword *\textit{the actual code}* is replaced by the code itself.

First, we naturally ask the model to search for the specific algorithm we are interested in.
We also ask the model to ignore functions whose code is not visible.
In this sense, we are concerned about hallucinations.
Because the model has been trained with large code bases, it could have been trained with the same code (at least, parts of the code) that we will be analyzing.
This can lead the model to think that it knows the code for unseen functions, which is actually wrong because supposing that a function with the same name and arguments corresponds to another function is no more than an assumption.

Next, we specify the output format.
We use ``Desired output'' to clearly indicate that follows the output format specification.
To make the output easy to parse, we ask the model to output ``Yes'' or ``Not'', followed by the name of the function (or functions) in case the answer is affirmative.
Afterward, we simply paste the code between three quotation marks to delimit the beginning and end of the source code.

\subsection{Second Prompt}
In the first prompt, we have clearly stated the task to perform.
Thus, we expect to have good detection results when the code contains any of the algorithms.
However, it is not clear how the prompt allows to discard false positives or reduce hallucination effects.
Hence, the second iteration of the prompt tries to accomplish this matter.

We considered using several prompt engineering techniques like chain-of-thought~\cite{cot,cot2}, self-consistency~\cite{selfconsistency} or tree-of-thoughts~\cite{tot}.
However, none of these techniques apply to code detection.
First, code detection does not follow any reasoning to conclude whether a code corresponds to a class of algorithms or not (it is more of a classification-like task).
Second, all of these techniques focus on improving the reasoning capabilities of LLMs in complex tasks, not mitigating hallucinations.
Previous works have highlighted that LLMs are able to identify when they have produced a wrong answer~\cite{shinn2023reflexion}.
This, however, requires several prompts.
The first one contains the task to perform by the LLM, and the second, where the LLM can use self-reflection to identify whether the previous answer is valid or not.
This two-step prompting of describing the task in a first prompt and realizing that it was wrong in a second prompt inspired us to propose a novel prompting technique for code detection.

\begin{figure}[ht!]
\caption{Second prompt (part 1).} \label{fig:prompt2}
\begin{mdframed}[style=simple3]
Can you explain what the following code does?\\
\\

Code:``````\\
*\textit{the actual code}* \\
''''''
\end{mdframed}
\end{figure}

\begin{figure}[ht!]
\caption{Second prompt (part 2).} \label{fig:prompt3}
\begin{mdframed}[style=simple3]
Does the code contain any function performing a *\textit{algorithm}*? Please ignore functions whose definition is not visible.\\
\\

Desired format:\\
Yes: {function name} (if there is a function).\\
No (if there is no function)\\
\end{mdframed}
\end{figure}

The second prompt is composed of two phases which are shown in Figures~\ref{fig:prompt2} and~\ref{fig:prompt3}.
Rather than following a zero-shot approach (like in the first prompt) here we use two prompts.
In the first part, we simply ask the model to explain what is the code doing.
Leaving the model to freely explain what the code does works very well because it is very easy for the model to understand what the task is.
Most of the time, the explanations given by the model at this step are correct (hallucinations are not present, or at least are very rare).
Once the model has analyzed the code, we ask, in a second prompt, if the code contains a given algorithm.

Please note the difference between this and the previous prompt.
Previously, we asked directly whether the code contains an algorithm.
Now, we ask the model to describe the code and then we ask if in that ``description'' that the model gave is found the algorithm in question.
In essence, false positives may arise in the first prompt (e.g., the model wrongly identifying parts of the code) but it is way less probable than asking directly to check for the algorithm.
In contrast, false positives may not appear as a consequence of the second prompt because the model is simply reusing the information previously given.
Thus, we expect to reduce hallucination effects with this technique, while maintaining similar detection results.

\subsection{Output Formatting} \label{sec:rules}
When the model's output matches exactly the expected output, the wrapper does not perform any parsing.
Here follows a description of the rules that the wrapper applies when the output does not match:

\begin{itemize}
\item \textbf{If the expected answer is positive, e.g.: ``Yes: (function list)''}: The wrapper removes the following substrings from the output: ``\textbackslash nNo'', ``\textbackslash n'', ``.'', ``()'' (where \textbackslash n is a new line). This makes it possible to accept outputs that contain outputs containing any ``garbage''.
\item \textbf{If the expected answer is negative, e.g.: ``No''}: The wrapper removes the following substrings, which we observed that occasionally appear: ``the code does not contain any function'', ``there is no function'', ``the code does not contain any function'' and ``there is no function''.
\end{itemize}

\section{Evaluation} \label{sec:s5}

\subsection{Setup} \label{sec:setup}
We evaluate the GPT-3.5 model using the source code dataset shown in Section~\ref{sec:dataset} with the two proposed prompts.
More precisely, we use \texttt{model=`gpt-3.5-turbo-16k'}, \texttt{temperature=0.0}, \texttt{top\_p=1.0} and \texttt{max\_tokens=512}.
The selected model allows inputs up to 16K tokens, allowing us to analyze larger codes, which would not be possible with the default GPT-3.5 model (which supports inputs with up to 4K tokens).
We aim to obtain answers as deterministic as possible, so we set \texttt{top\_p} to 1.0 (the default value) and only modify \texttt{temperature}, which we set to 0.
Lastly, \texttt{max\_tokens} sets the maximum number of allowed tokens in the output, which we limit to 512 since it is enough for the first step of the second prompt.

In programs with multiple valid functions (e.g., multiple functions performing a GEMM), we expect the model to find the outermost function.
For each prompt, we show the confusion matrix and also a summary matrix that presents how each algorithm is classified.
In the false positives evaluation for a given algorithm, we also include real programs from other algorithms (e.g., for GEMM we add convolution and FFT codes to the false positives dataset).
Besides, we provide a detailed explanation of why the model was unable to find the algorithm.
We identify three types of errors: 
\begin{itemize}
\item Error 1: GPT-3.5 thinks there is no function, where there is actually at least one.
\item Error 2: GPT-3.5 finds at least one function, but not the one we are looking for (the outermost).
\item Error 3: Wrong output format (the output is right, but the Python wrapper is not able to parse it).
\end{itemize}

\subsection{First Prompt}
Table~\ref{tab:matrix1} shows the confusion matrix for the three analyzed algorithms.
True positive results are excellent in all cases, achieving 93\% in GEMM and 100\% in convolution and FFT codes.
This seems to indicate that the model can confidently analyze the code and find if the algorithm is present or not.
Conversely, the number of false positives is exceedingly high in the three algorithms, and it is even more notable in the case of GEMM and convolution.
Alternatively, we can compute the accuracy of the model as:

\begin{align*}
\text{Accuracy} = \frac{TP+TN}{TP+TN+FP+FN}
\end{align*}

which, using data from Table~\ref{tab:matrix1}, yields a poor accuracy for Convolution (22.3\%), followed by GEMM (68.8\%) and FFT (79.2\%).
Despite the high precision, the accuracy is severely harmed due to the high false positive rate.
Results seem to indicate that the model has a high tendency to answer our questions affirmatively, rather than reasoning about the code and answering accordingly.

\begin{table}[t]
\caption{Confusion matrices for GEMM, CONV and FFT (first prompt).} \label{tab:matrix1}

\centering

\bgroup
\def\arraystretch{2.25} %

\begin{minipage}{0.35\linewidth}
\setlength\tabcolsep{2.0pt} %
\begin{tabular}{l|l|c|c|c}
\multicolumn{2}{c}{}  & \multicolumn{2}{c}{} & \\
\hhline{~~|-|-|}
\multicolumn{2}{c|}{} & T & N \\
\hhline{~|-|-|-|}
\multirow{2}{*}{\rotatebox[origin=c]{90}{Actual}\hspace*{0.15cm}} & T & \confmatl{93}{119}{white} & \confmatr{7}{9}{black} \\
\hhline{~|-|-|-|}
& N & \confmatl{43.2}{111}{black} & \confmatr{56.8}{146}{white} \\
\hhline{~|-|-|-|}
\multicolumn{2}{c}{}  & \multicolumn{2}{c}{GEMM} \\
\end{tabular}
\end{minipage}
\hspace*{0.05cm}
\begin{minipage}{0.30\linewidth}
\setlength\tabcolsep{2.0pt}
\begin{tabular}{|c|c|}
\multicolumn{2}{c}{Predicted} \\
\hline
T & N \\
\hline
\confmat{100}{15}{white} & \confmatr{0}{0}{black} \\
\hline
\confmat{80.8}{299}{white} & \confmatr{19.2}{71}{black} \\
\hline
\multicolumn{2}{c}{CONV} \\
\end{tabular}
\end{minipage}
\hspace*{-0.40cm}
\begin{minipage}{0.30\linewidth}
\setlength\tabcolsep{2.0pt}
\begin{tabular}{|c|c|}
\multicolumn{2}{c}{} \\
\hline
T & N \\
\hline
\confmat{100}{15}{white} & \confmatr{0}{0}{black} \\
\hline
\confmat{21.6}{80}{black} & \confmatr{78.4}{290}{white} \\
\hline
\multicolumn{2}{c}{FFT} \\
\end{tabular}
\end{minipage}

\egroup

\end{table}

Regarding output consistency according to the rules set in the prompt, we found that the rules implemented in the wrapper (described in Section~\ref{sec:rules}) are rarely triggered.
For the case where the expected output is affirmative, the most simple rules like removing a dot are triggered, but not very often.
In fact, when the output is negative, rules never get triggered.
This indicates that the output formatting rules in the prompt work consistently well since very little parsing is needed.

Table~\ref{tab:errors1} presents the type of false negatives for each algorithm.
The most common case is caused by the model not giving a valid output format.
Here we mostly found issues with C++ formatting, where our Python wrapper was expecting to find the name of the function, but the model also includes C++ artifacts.
The model answered a valid function, but not the outermost function in two cases, while the rest correspond to the model simply answering that there were no functions matching the algorithm.

\begin{table}[ht]
\caption{Summary of false negatives types (first prompt).} \label{tab:errors1}
\centering
\begin{tabular}{llll}
\toprule
        & GEMM & CONV & FFT \\ \midrule
Error 1 & 2    & 0    & 0   \\
Error 2 & 2    & 0    & 0   \\
Error 3 & 5    & 0    & 0   \\ \midrule
Total   & 9    & 0    & 0   \\ \bottomrule
\end{tabular}
\end{table}

Regarding false positives, results show that the model gets confused and identifies algorithms where they are not.
It is worth noticing that these false positives always occur on functions that are defined and declared in the code.
The only exception to this rule is found in Caffe, where we found that the model also triggers false positives in functions that contain ``gemm'' in the function name, even if they are not defined or declared.
It seems reasonable to think that such functions perform matrix multiplication, but if code is not available it is only an assumption that the model is unable to confirm.
This appears to be a clear example of external hallucination since the model thinks to know the content of the function, which is actually unavailable.
We also observed clear cases of hallucination in some programs where there was no function to detect but the model gave an output with a length equal to \texttt{max\_tokens} (e.g., exhausting the output) repeating one function name over and over.

\begin{figure}[ht]
\centering
\includegraphics[width=0.6\linewidth]{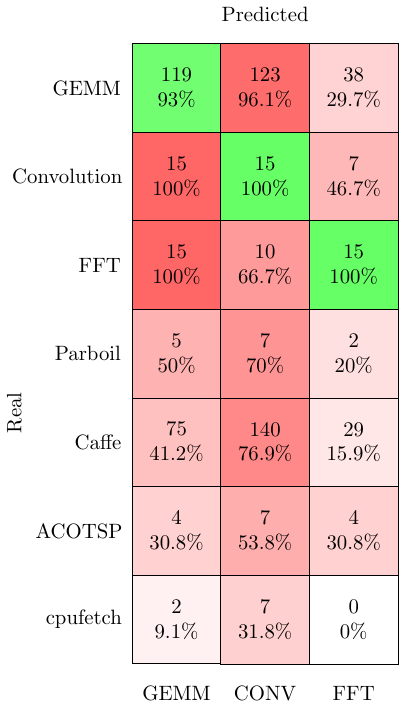}
\caption{Summary of confusion matrix (first prompt).} \label{fig:summary1}
\end{figure}

Figure~\ref{fig:summary1} depicts the results summary for the first prompt, showing for each algorithm (GEMM, CONV, FFT) in the x-axis, the percentage of matched programs against each code type (in the y-axis).
Ideally, we would like to have as highest values as possible for the diagonal of the first three elements, while the rest of the cells are as close to zero as possible.
Having high values in the diagonal indicates high true positive rates, while low values in the rest of the figure indicate low false positives.
In the first prompt, we find high values in the diagonal, as well as in the rest of the figure.
The highest false positive rates are found when we ask the model if a GEMM program contains any convolution (96.1\%), and when we ask if convolution or FFT programs contain GEMMs (100\% in both cases).
Results evidence that the model gets easily confused when mixing GEMM and convolution (asking for GEMM in convolution programs and vice versa).
This might be motivated by the fact that both have relatively similar code structures and that sometimes convolutions are implemented with matrix multiplications.
However, we also find surprisingly high false positives when analyzing other codes.
In the Parboil benchmark, we obtain rates as high as 50\% and 70\% for GEMM and convolution, respectively.
This benchmark suite contains programs for evaluating the performance of microprocessors, which also have certain similarities with GEMM and convolution.
In Caffe, the false positive rate is also high.
The model gets easily confused with this code because many of them have functions containing ``gemm'' or ``conv'' in the code.
However, most of the time they are simply function calls, rather than function definitions.
It is surprising to find that the model gets confused with these even though we explicitly asked to ignore functions whose definition is not visible.
Also, sometimes functions are visible, but they do not perform GEMM or convolutions explicitly in the code.
Lastly, we also found surprisingly high false positive values in ACOTSP-MF and cpufetch, even though those programs have in no way any similar code to GEMM or convolution.

\subsection{Second Prompt}
Table~\ref{tab:matrix2} shows the confusion matrix using the second prompt.
The first thing that draws attention is the true positives.
They are very similar to the one we had in the first prompt, but they are slightly lower.
However, the number of false positives has decreased significantly, with only 5.8\%, 1.6\% and 0.0\% for GEMM, convolution and FFT, respectively.
These results confirm that our prompt proposal drastically improves the accuracy of the model, which raises to 91.1\%, 97.9\% and 99.7\%, respectively.

\begin{table}[ht]
\caption{Confusion matrices for GEMM, CONV and FFT (second prompt).} \label{tab:matrix2}
\centering

\bgroup
\def\arraystretch{2.25}

\begin{minipage}{0.35\linewidth}
\setlength\tabcolsep{2.0pt}
\begin{tabular}{l|l|c|c|c}
\multicolumn{2}{c}{}  & \multicolumn{2}{c}{} & \\
\hhline{~~|-|-|}
\multicolumn{2}{c|}{} & T & N \\
\hhline{~|-|-|-|}
\multirow{2}{*}{\rotatebox[origin=c]{90}{Actual}\hspace*{0.15cm}} & T & \confmatl{85.2}{109}{white} & \confmatr{14.8}{19}{black} \\
\hhline{~|-|-|-|}
& N & \confmatl{5.8}{15}{black} & \confmatr{94.2}{242}{white} \\
\hhline{~|-|-|-|}
\multicolumn{2}{c}{}  & \multicolumn{2}{c}{GEMM} \\
\end{tabular}
\end{minipage}
\hspace*{0.05cm}
\begin{minipage}{0.30\linewidth}
\setlength\tabcolsep{2.0pt}
\begin{tabular}{|c|c|}
\multicolumn{2}{c}{Predicted} \\
\hline
T & N \\
\hline
\confmat{86.7}{13}{white} & \confmatr{13.3}{2}{black} \\
\hline
\confmat{1.6}{6}{black} & \confmatr{98.4}{364}{white} \\
\hline
\multicolumn{2}{c}{CONV} \\
\end{tabular}
\end{minipage}
\hspace*{-0.40cm}
\begin{minipage}{0.30\linewidth}
\setlength\tabcolsep{2.0pt}
\begin{tabular}{|c|c|}
\multicolumn{2}{c}{} \\
\hline
T & N \\
\hline
\confmat{93.3}{14}{white} & \confmatr{6.7}{1}{black} \\
\hline
\confmat{0}{0}{black} & \confmatr{100}{370}{white} \\
\hline
\multicolumn{2}{c}{FFT} \\
\end{tabular}
\end{minipage}

\egroup

\end{table}

In the second prompt, we found that the rules implemented in the wrapper (shown in Section~\ref{sec:rules}) are much more likely to be triggered than in the first prompt.
Specifically, rules for the case where the expected output is negative were never triggered in the first prompt, but they are sometimes triggered in this prompt.
These results indicate that the output specification is less robust in the second prompt compared to the first one which showed to be pretty reliable.

\begin{table}[ht]
\caption{Summary of false negatives types (second prompt).} \label{tab:errors2}
\centering
\begin{tabular}{llll}
\toprule
        & GEMM & CONV & FFT \\ \midrule
Error 1 & 1    & 1    & 0   \\
Error 2 & 11   & 1    & 1   \\
Error 3 & 7    & 0    & 0   \\ \midrule
Total   & 19   & 2    & 1   \\ \bottomrule
\end{tabular}
\end{table}

\begin{figure}[ht]
\centering
\includegraphics[width=0.6\linewidth]{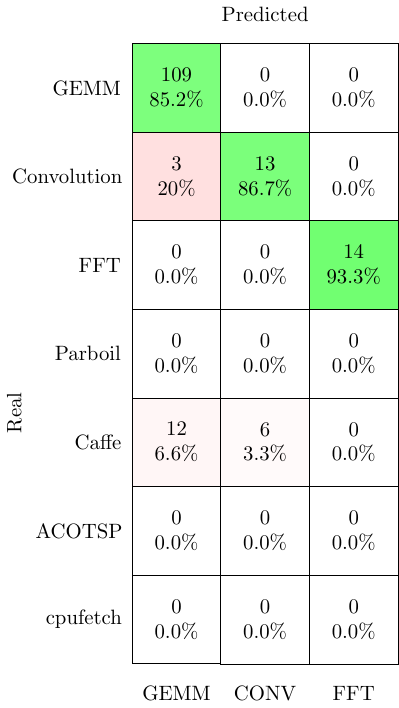}
\caption{Summary of confusion matrix (second prompt).} \label{fig:summary2}
\end{figure}

We analyze the failure reasons for the second prompt in Table~\ref{tab:errors2}.
The majority of failures are found in Error 2, (e.g., when the model finds at least one function performing a matrix multiplication, but it is not the one we were expecting to find).
Besides, we still have the same problems in GEMM programs where the model does not comply with the output format that we expect, which makes the wrapper unable to parse the output.

In the second prompt, false positives only arise when looking for GEMM and convolution in Caffe, and when looking for GEMM in convolution codes.
As we mentioned in the first prompt, the model had two types of false positives in Caffe.
The majority were functions declared and defined in the code, while a minority were undefined functions containing ``gemm'' in the name.
In the second prompt, all the false positives from Caffe correspond to the second class.
This indicates that the model still hallucinates with functions that have ``gemm'' in its name, while the other false negatives have completely disappeared.
Lastly, in convolution codes, we can still find some cases where the code is purely performing a convolution but the model reports it as performing a matrix multiplication.
Sometimes, it happens because the code contains a matrix multiplication inside the reported function, but it does not mean that the function performs matrix multiplication exclusively (which is what we are asking).
However, we believe that these issues should be fixed with more powerful models (e.g., GPT-4) or code-specific models (e.g., AlphaCode)

\section{Conclusions and Future Work} \label{sec:s6}
Large Language Model's scale and complexity have grown massively in the last few years.
ChatGPT in particular and LLMs in general have caused a massive impact on society due to the enormous potential of these models for greatly diverse natural language tasks.
We also expect those models to keep scaling and improving in the near future, so finding new applications where these models excel are key for fully exploiting them.
A field not previously explored with LLMs was code detection, which is key for many applications in programming language research.
Particularly, it has been studied to achieve code lifting, a technique that consists in replacing handwritten code with a call to an optimized library.
Previous work has approached this topic either with constraint-based matching or with neural embeddings plus input/output equivalence, which can be costly for large codes.

In this work, we have explored the application of LLMs to code detection for the first time.
Specifically, we evaluated GPT-3.5 in code detection with matrix multiplication (GEMM), convolution and fast-fourier transform (FFT) algorithms.
After designing our first prompt for code detection, the model showed an accuracy of 68.8\%, 22.3\% and 79.2\% for GEMM, convolution and FFT, respectively.
False positives, triggered by hallucinations, are the reason to explain such poor results.
In the second prompt, we introduced a novel approach for code detection that achieved an accuracy of 91.1\%, 97.9\% and 99.7\%, respectively.
The new prompt drastically reduces the number of false positives which, still occurring, are way less frequent.
Despite not being trained specifically for code detection, GPT-3.5 results are truly impressive, reaching an accuracy very close to 100\%.

Rather than using raw code input, we aim to explore alternative approaches in future research.
Instead of analyzing program files individually, we are interested in developing a novel methodology that focuses on analyzing the program function by function.
This approach would involve creating a dataflow graph that captures the interconnections between the functions defined in the program.
Adopting this approach would allow us to analyze the entire code structure as opposed to processing one file at a time.
Additionally, we are intrigued by the performance of other models like GPT-4 or other LLMs trained specifically on source code.
We anticipate that these models will achieve even better results, pushing the limits further towards achieving 100\% accuracy.

\section*{Acknowledgements}
Grant TED2021-129221B-I00 funded by MCIN/AEI/10.13039/501100011033 and by the ``European Union NextGenerationEU/PRTR''.

\bibliographystyle{plain}
\bibliography{refs}

\end{document}